\documentclass[3p,12pt]{elsarticle}
\usepackage{eurosym}
\usepackage{amssymb}
\usepackage{float,epsfig,latexsym,dcolumn,lscape,longtable,setspace}
\usepackage{makeidx}
\usepackage{graphicx}
\usepackage{amsmath}
\usepackage{fancyhdr}
\usepackage{subfigure}
\usepackage{threeparttable}
\usepackage{epstopdf}
\usepackage{booktabs}
\usepackage[expansion=true,protrusion=true]{microtype}
\usepackage{amsmath}
\usepackage[T1]{fontenc}
\usepackage[sc]{mathpazo}
\usepackage{color}
\usepackage[colorlinks=true]{hyperref}
\usepackage{mathtools}
\usepackage{rotating}
\usepackage{afterpage}
\usepackage{tabularx}

\definecolor{mydarkblue}{rgb}{0,0,0.3}
\AtBeginDocument{\hypersetup{linkcolor=mydarkblue,citecolor=mydarkblue}}

\makeatletter
\def\ps@pprintTitle{  \let\@oddhead\@empty
  \let\@evenhead\@empty
  \let\@oddfoot\@empty
  \let\@evenfoot\@oddfoot
}
\makeatother

\newcolumntype{d}[1]{D{.}{.}{#1}}
\newcolumntype{p}[1]{D{(}{(}{-1}}

\begin{document}

\bigskip
\begin{frontmatter}

\title{The Price of BitCoin:\\
GARCH Evidence from High Frequency Data\tnoteref{thanks}}

\tnotetext[thanks]{The authors gratefully acknowledge financial support received from the Slovak Research and Development Agency under the contract No. APVV-15-0552 and VEGA 1/0797/16. The authors would like to thank Gerald Dwyer as well as participants of the Econometric Society, the International Conference on Macroeconomic Analysis and International Finance in Creta as well as seminar participants at the European Commission for comments and useful suggestions. The authors are solely responsible for the content of the paper. This paper is based on the JRC study 'The Price of BitCoin: GARCH Evidence from High Frequency Data'. The views expressed are purely those of the authors and may not in any circumstances be regarded as stating an official position of the European Commission.}

\author[JRC,LICOS]{Pavel Ciaian}
\author[JRC,LICOS]{d'Artis Kancs\corref{dartis}}
\ead{d'artis.kancs@ec.europa.eu}
\cortext[dartis]{Corresponding author, Competence Centre on Modelling, European Commission.}
\author[SAU,UWB]{Miroslava Rajcaniova}

\address[JRC]{European Commission, DG Joint Research Centre, I-21027 Ispra, Italy.}
\address[LICOS]{LICOS, University of Leuven, B-3000 Leuven, Belgium}
\address[SAU]{SAU, Department of Economic Policy SK-94901 Nitra, Slovakia}
\address[UWB]{University of West Bohemia, CZ-30614 Pilsen, Czech Republic}

\begin{abstract}

This is the first paper that estimates the price determinants of BitCoin in a Generalised Autoregressive Conditional Heteroscedasticity framework using high frequency data. Derived from a theoretical model, we estimate BitCoin transaction demand and speculative demand equations in a GARCH framework using hourly data for the period 2013-2018. In line with the theoretical model, our empirical results confirm that both the BitCoin transaction demand and speculative demand have a statistically significant impact on the BitCoin price formation. The BitCoin price responds negatively to the BitCoin velocity, whereas positive shocks to the BitCoin stock, interest rate and the size of the BitCoin economy exercise an upward pressure on the BitCoin price.

\end{abstract}


\end{frontmatter}

\vfill

\begin{center}
\today
\end{center}

\renewcommand{\thefootnote}{\arabic{footnote}} \setcounter{footnote}{0} %
\thispagestyle{empty}\addtocounter{page}{-1} \onehalfspacing
\newpage

\section{Introduction}

During the last decade, the rise of virtual currencies has triggered a
growing interest in the economic literature -- both theoretical models and
empirical studies have attempted to understand drivers behind growth and the
mechanics of virtual currencies. The large majority of the existing
empirical literature on virtual currencies is based on rather aggregated
(either on daily or weekly) data though, which masks a great deal of
complexity surrounding, for example, the virtual currency price formation.

Previous studies have looked at various factors related to the blockchain
technology and its implication for financial markets (e.g. Grinberg 2011;
Barber et al. 2012; Moore and Christin 2013; Bouri et al. 2017; Baur et al.
2018; Gandal et al. 2018). A frequently analysed issue in literature relates
to the understanding of the virtual currency price formation (e.g. Buchholz
et al. 2012; Kristoufek 2013; van Wijk 2013; Bouoiyour and Selmi 2015;
Bouoiyour et al. 2016; Ciaian et al. 2016, 2018; Aalborg et al. 2018; Jang
and Lee 2018). Several determinants of virtual currency prices have been
identified as important in the previous literature, such as market forces of
supply and demand (Buchholz et al. 2012; Bouoiyour and Selmi 2015; Aalborg
et al. 2018; Baur et al. 2018; Jang and Lee 2018), speculations (Kristoufek
2013; Bouoiyour and Selmi 2015) and macro-financial developments (van Wijk
2013 Ciaian et al. 2016, 2018).

Despite the growing literature in this field, the existing evidence is still
rather inconclusive in terms of providing a conceptually and empirically
consistent explanation of the BitCoin price dynamics. The present study
attempts to shed an additional light on this highly complex BitCoin price
dynamics by making use of high frequency data. To our knowledge, this is the
first paper that estimates the price determinants of BitCoin in a
Generalised Autoregressive Conditional Heteroscedasticity (GARCH) framework
using high frequency data.

In order to identify and estimate drivers of the BitCoin price, first we
derive a conceptual model of the BitCoin price formation (section 2).
Building on Mankiw (2007) and Ciaian et al. (2016), in the present study we
rely on a conceptual framework which considers both the transaction demand
and speculative demand for money (store of value) in order to understand the
mechanics behind the BitCoin price formation,

In a second step, building on previous empirical studies on the BitCoin
price formation, we specify a GARCH model to estimate factors affecting the
BitCoin price using hourly data for the period 2013--2018 (sections 3 and 4,
respectively). The necessity to depart from traditional time-series
analytical mechanisms is given by the fact that virtual currencies are
highly volatile compared to traditional currencies. As a result, their
exchange rates cannot be assumed to be independently and identically
distributed. Given that this virtual currency property violates the
assumption of a constant conditional variance given past information
(required e.g. in ARMA models), usually, empirically studies (Chen et al.,
2016; Bouoiyour and Selmi, 2015 and 2016; Cermak, 2017; Dyhrberg, 2016a and
2016b and others) choose the GARCH approach of Engle (1982) and Bollerslev
(1986) to model the historical volatility of virtual currency prices.
According to Kalev et al. (2004), the modelling of volatility through a
conditional heteroscedasticity process presents a great improvement over
unconditional volatility models.

Our empirical results confirm that both the BitCoin transaction demand and
speculative demand have a statistically significant impact on the BitCoin
price formation (section 5). The BitCoin price responds negatively to the
BitCoin velocity, whereas positive shocks to the BitCoin stock, interest
rate and the size of the BitCoin economy exercise an upward pressure on the
BitCoin price. The high frequency (hourly) data analysed in the present
study allow to gain additional insights, which remain masked using averaged
daily or weekly prices. To our knowledge, this is the first study in
literate using high frequency data in the context of the BitCoin price
analysis.

\section{Conceptual framework}

\subsection{BitCoin versus standard currencies}

Similar to standard currencies, the BitCoin economy regulates the total
money supply by adjusting both the stock of money in circulation and its
growth rate. However, neither the stock nor the growth rate of money is
controlled by a centralised financial authority or government, but instead
by a software algorithm. Both are exogenously pre-defined and publicly known
to all market participants from the time of the BitCoin launch. This BitCoin
feature contrasts standard currencies, where the supply of money is at the
discretion of Central Banks and thus not known a-priori (i.e. it depends on
macroeconomic developments and the monetary policy of the Central Bank).
This implies that the BitCoin money supply is exogenous.

The BitCoin demand (in dollar denomination) depends on the transaction
demand and speculative demand. The transaction demand for money/currency
arises from the absence of a perfect synchronisation of payments and
receipts. Market participants may hold money/currency to bridge the gap
between payments and receipts and to facilitate daily transactions. BitCoin
has several advantages which may make it the preferred choice for being used
as a medium of exchange. Among others, BitCoin advantages include the
relatively fast transaction execution, the relatively low transaction fees
and a certain level of anonymity given that BitCoin transactions are
nameless and do not require the provision of personal identity information.

The speculative (investment) demand stems from potential profit
opportunities that may arise on financial markets and refers to cash held
for the purpose of avoiding a capital loss from investments in financial
assets, such as bonds. A rise in the financial asset return (e.g. interest
rate) causes their prices to fall, leading to a capital loss (negative
return) from holding financial assets. Thus, investors may prefer to hold
money/BitCoin to avoid losses from financial assets (Keynes, 1936). This
implies a negative relationship between virtual currencies and the interest
rate. The BitCoin demand for speculative purposes could be driven by its use
as a safeguarding against inflation or financial market uncertainties. In
such situations, BitCoin might be the preferred option for holding it for
the purpose to avoid capital losses from holding financial assets (e.g.
bonds) (Folkinshteyn et al. 2015; Baur et al. 2018; Ciaian et al. 2018).

\subsection{The model}

According to Mankiw (2007), the transaction demand and speculative demand
for money are the key factors affecting any currency's price formation. In
the context of BitCoin, Ciaian et al. (2016, 2018) and Baur et al. (2018)
show that the price formation of BitCoin can be studied by considering the
interaction between the supply and demand drivers of the BitCoin economy
(e.g. the amount of coins in circulation and transaction/speculative
demand). Building on Mankiw (2007) and Ciaian et al. (2016), in the present
study we rely on a conceptual framework which considers both the transaction
demand and speculative demand for money (store of value) in order to
understand the mechanics behind the BitCoin price formation.

Given that the BitCoin money supply, $M^{S}$, is exogenous (see section 2.1)
in terms of a standard currency, it can be expressed as a product of the
total stock of BitCoin in circulation, $B$, and the exchange rate of the
virtual currency (i.e. dollar per unit of virtual currency), $P$:

\begin{equation}
M^{S}=PB
\end{equation}

BitCoin transactions can be executed between decentralised agents. Similar
to standard currencies, BitCoin can be used as a medium of exchange
(transaction demand for money) and a store of value (speculative demand for
money). However, unlike standard currencies, there are no physical coins
linked to BitCoin transactions. Instead, there are \textquotedblleft digital
BitCoins\textquotedblright\ that are stored digitally on a global database
(or blockchain). Blockchain records all transactions and is checked and
validated by a peer-to-peer network of computers around the world.

The transaction demand for money, $M^{D}$, can be defined as a constant
proportion k of the size of the BitCoin economy, $G$, while the speculative
demand for money is a function of the interest rate, $L(i)$, with $\partial
L/\partial r<0$. Variable $1/k$ represents the velocity of BitCoin in
circulation, whereas $G$ approximates the volume of transactions (Howden
2013):

\begin{equation}
M^{D}=kG+L\left( i\right)
\end{equation}

In equilibrium, the BitCoin supply (1) and demand (2) price relationship can
be expressed as:

\begin{equation}
P=\frac{kG+L\left( i\right) }{B}
\end{equation}%
According to equation (3), the price of BitCoin decreases with velocity, the
BitCoin stock and interest rate, but increases with the size of the BitCoin
economy. Equation (3) implies that the price of BitCoin can be sustained
only if market participants use it as medium to intermediate the exchange of
goods and services or as a store of value (i.e. for speculative purposes).

\section{Econometric approach}

\subsection{Previous studies}

Gronwald (2014) was among first who estimated an autoregressive
jump-intensity GARCH model in the context of virtual currencies. He finds
that BitCoin prices are strongly characterised by extreme price movements,
which is an indication of an immature market.

Bouoiyour and Selmi (2015) investigated daily BitCoin prices using a variety
of GARCH models. They find that volatility has decreased when comparing data
from 2010--2015 with data from the first half of 2015. During the first time
interval, threshold GARCH estimates revealed a great duration of
persistence. In the second period, exponential GARCH results displayed less
volatility persistence. In a follow-up study, Bouoiyour and Selmi (2016)
applied several GARCH extensions, such as the exponential GARCH, the
asymmetric power ARCH, the weighted GARCH and multiple threshold-GARCH
specifications. Their results suggest that, despite maintaining a moderate
volatility, BitCoin remains typically reactive to negative rather than
positive news and BitCoin market is therefore, still far from being mature.

Letra (2016) used a GARCH (1,1) specification to analyse daily BitCoin
prices and search trends on Google, Wikipedia and tweets on Twitter. He
found out that BitCoin prices were influenced by popularity, though also
that the web content and BitCoin prices are connected; they exhibit certain
predictable power.

Chen et al. (2016) used various specifications of GARCH models to analyse
the CRIX index family using daily data from 2014--2016. The authors conclude
that the TGARCH (1,1) model is the best fitting model for all sample data
based on discrimination criteria of loglikelihood, AIC and BIC.

Dyhrberg (2016a) applied the asymmetric GARCH methodology to explore the
hedging capabilities of BitCoin. Dyhrberg concludes that BitCoin can be used
as a hedge against stocks in the Financial Times Stock Exchange Index and
against the USA dollar in the short term. In a related work, Dyhrberg
(2016b) used GARCH models to explore the financial asset capabilities of
BitCoin. Results suggest that BitCoin has a place on the financial markets
and in portfolio management, as it can be classified as something in between
gold and the USA dollar, on a scale from a pure medium of exchange to a pure
store of value.

Cermak (2017) used a GARCH (1,1) specification to model the BitCoin's
volatility with respect to macroeconomic variables, in countries where
BitCoin is being traded the most. The results show that the macroeconomic
explanatory variables of China, the United States, and the EU are
significant to forecast the next day's volatility of BitCoin. BitCoin is
starting to react to the same variables as the fiat currencies in these
countries. Japan's macroeconomic variables are not significant, however.

Chu et al. (2017) used GARCH to model seven most popular virtual currencies.
Their results suggest that virtual currencies such as BitCoin, Ethereum,
Litecoin and many others display a relatively high volatility, especially at
their inter-daily prices. Chu et al. (2017) conclude that this type of
investment is suited for risk-seeking investors looking for a way to invest
or enter into technology markets.

Urquhart (2017) examined BitCoin's volatility and the forecasting ability of
GARCH and HAR models in the BitCoin market. He finds that the realised
volatility is quite high in the first half of the sample but has decreased
in the recent years -- a finding that is consistent with Bouoiyour and Selmi
(2015).

Stavroyiannis and Babalos (2017) investigated the dynamic properties of
BitCoin modelling through univariate and multivariate GARCH models and
vector autoregressive specifications. They concluded that BitCoin does not
actually hold any of the hedge, diversifier or safe-haven properties;
rather. Instead, it exhibits intrinsic attributes not related to USA market
developments.

Katsiampa (2017) estimated the volatility of BitCoin through a comparison of
GARCH models. A GARCH model with an AR transformation fitted daily data
best, which emphasis the significance of including both the short and long
run component of the conditional variance (Katsiampa, 2017).

\subsection{Empirical specification}

Based on the theoretical model outlined in section 2, we derive an
econometrically estimable BitCoin price (return) equation. The (Generalised)
Autoregressive Conditional Heteroscedasticity (G)ARCH approach adopted in
the present study is particularly suited for capturing the volatility
clustering that is characteristic for financial time series (including
BitCoin) where typically data show continuous periods of a high volatility
and continuous periods of a low volatility. Following the exchange rate
volatility literature (Poon, Granger, 2005; Hansen and Lunde, 2005;
Brownlees et al., 2011), in the present study we model the historical
volatility of BitCoin prices by specifying a GARCH model. GARCH takes into
account the excess kurtosis (i.e. fat tail behaviour) and volatility
clustering -- two important characteristics of financial time series, which
are also observable in the Bitcoin case.

Let $r_{t}$ denote the log returns of BitCoin prices:

\begin{equation}
r_{t}=\ln \left( P_{t}\right) -\ln \left( P_{t-1}\right)
\end{equation}%
where $r_{t}$ are log returns at time $t$; Pt denotes the price of BTC in
USD at time $t$. A GARCH(1,1) model can then be specified as:

\begin{equation}
r_{t}=\mu _{t}+\sigma _{t}\varepsilon _{t}
\end{equation}%
where $\mu _{t}$ is the conditional mean; $\sigma _{t}$ is the volatility
process; $\varepsilon _{t}$ denotes residuals of the volatility.

As usual, we start with specifying a conditional mean equation, which is
assumed to be an AR(1) process, implying that the returns in the previous
period are used to predict the returns of the current period:

\begin{equation}
r_{t}=\beta _{0}+\beta _{1}r_{t-1}+\varepsilon _{t}\text{ \ \ \ with \ \ \ }%
\varepsilon _{t}\approx i.i.d.\left( \theta ,\sigma ^{2}\right) \text{ \ \ \
\ and \ \ \ \ }\mid \theta \mid <1
\end{equation}%
Residuals of the estimated mean equation are then tested for the presence of
ARCH effects using the Lagrange multiplier (LM) test for autoregressive
conditional heteroscedasticity in residuals.

The variance of the dependent variable is modelled as a function of the past
values of the dependent variable and independent or exogenous variables. The
GARCH framework allows variance not only to depend on past shocks but also
to depend on the most recent variance of itself. The specification for the
conditional variance of GARCH(q,p) follows Bollerslev (1986) and can be
represented as follows:\footnote{%
We follow the standard notation in literature, whereby a GARCH model of
order "p" and "q" or GARCH(p,q) indicates the number of lags of the squared
residual return ("p") and the number of lags of variances ("q") included in
the model.}

\begin{equation}
\sigma _{t}^{2}=\omega +\sum_{t=1}^{p}a_{i}\varepsilon
_{t-i_{1}}^{2}\sum_{j=1}^{q}\beta _{j}\sigma _{t-j}^{2}
\end{equation}%
where $\sigma _{t}^{2}$ is the conditional variance period $t$; $\omega $ is
the weighted long run average variance; $\varepsilon _{t-i_{1}}^{2}$ is the
squared residual return in the previous period (ARCH term); $\sigma
_{t-j}^{2}$ is the variance in the previous period (GARCH term); $%
a_{i}+\beta _{j}<1$ is the stationarity condition; and $\omega >0,$ $%
a_{i}>0, $ $\beta _{j}>0$ are a GARCH parameter restrictions.

According to equation (7), the conditional variance is a function of three
terms: (i) a constant term, $\omega $ (ii) news about volatility from the
previous period, measured as the lag of the squared residual from the mean
equation, $\varepsilon _{t-i_{1}}^{2}$ (the ARCH term); and (iii) the last
period's forecast variance, $\sigma _{t-j}^{2}$ (the GARCH term). The key
feature in a GARCH model is the sum of $\alpha $ and $\beta $, indicating
for how long volatilities persist after a price shock.

In order to investigate the impact of explanatory variables on the BitCoin's
volatility that have been identified in previous studies as important, both
AR(1) and GARCH(1,1) models are extended by exogenous explanatory variables
in conditional mean and conditional variance equations following Vlastakis
and Markellos (2012).

\section{Results}

The mean and variance equations of the GARCH model are estimated in five
different sets of alternative specifications, in order to account for
potential cross-correlations between variables and for the fact that we use
two alternative proxies for several variables (e.g. the BitCoin velocity).
The five alternative GARCH specifications follow closely Ciaian et al.
(2016), their differences are summarised in Table 2. Models 1.1--1.2 (Models
2.1--2.2) consider interchangeably BitCoin volume (\textit{logvolume}) and
the number of BitCoin users (\textit{logno}) and velocity (\textit{%
logvelocity}), while Models 1.3--1.5 (Models 2.3--2.5) consider
interchangeably the total BitCoin stock (\textit{logtot\_btc}), BitCoin
volume (\textit{logvolume}), the number of BitCoin users (\textit{logno})
and the two velocity variables (\textit{logvelocity, logvelocity2}).

\subsection{Data}

In the present study, we use hourly data for the period 2013--2018 with more
than 50 thousand observations in total. As a response variable, we use the
log returns of the daily Bitcoin price. The usage of log returns is well
documented in the empirical finance literature because most prices of
financial series are non-stationary. In the context of the present study, an
advantage of using log returns is that the data is normalised and normally
distributed. The log returns are defined as the first difference of the
natural logarithm of the prices (see equation (4) in section 3).

As regards explanatory variables, we measure the BitCoin economy, $G$, by
the volume of transactions, which can be further decomposed into the number
of transactions and the value per transaction. Given that $G$ is measured by
the volume of BitCoin transactions, $1/k$ can be represented by the velocity
of BitCoin circulation (Howden 2013). Note that velocity is an unobserved
variable. The fungibles of BitCoin implies that the frequency at which the
same BitCoin is used to purchase goods and services within a given time span
cannot not be straightforwardly tracked. Therefore, we proxy velocity by two
alternative variables computed by summing up all BitCoin transactions
processed on blockchain in every hour divided by the network's average
BitCoin base. These data are extracted from \textit{data.bitcoinity.org}.
The total stock of BitCoin in circulation, $B$, is calculated using daily
data on the total stock of BitCoin from \textit{quandl.com} and the average
time to mine a BitCoin block in minutes from \textit{data.bitcoinity.org}.
The total stock of BitCoin has been first calculated per minute and then
aggregated by hour. To account for the speculative demand for BitCoin, we
proxy for the interest rate, i, using the 10 Year Treasury Inflation Indexed
Security (daily) extracted from Federal Reserve Bank of St. Louis.\footnote{%
\url{https://research.stlouisfed.org/useraccount/datalists/202281/download}}

\subsection{Specification tests}

As usual, before modelling time series data we check for stationarity, as
the underlying econometric methodology is inherently based on the
stationarity assumption. According to results of the Augmented
Dickey--Fuller (ADF) tests, we fail to accept the null hypothesis of a unit
root for BitCoin returns and, hence, stationarity is guaranteed for the
log-return series of BitCoin prices.

As next, the statistical properties of the mean equation are examined. In
particular, two preconditions have to be met for a GARCH model: a clustering
volatility and a serial correlation of the heteroscedasticity.

In order to establish whether there is a clustering volatility in residuals,
first we plot them graphically. We can observe that periods of high
volatility are followed by periods of high volatility, while periods of low
volatility seem to be followed by periods of low volatility. This indicates
that in our series large returns are followed by large returns and small
returns are followed by small returns. This provides a clear indication of a
clustering of volatility, which means that the residual is conditionally
heteroscedastic. Similarly, excess kurtosis and fat tails characterising our
series suggest that the error term is conditionally heteroscedastic and can
be represented by a GARCH model.

The second criteria required for GARCH is a serial correlation of
heteroscedasticity. In order to determine whether there is a serial
correlation of the heteroscedasticity, we conduct the Engle's Lagrange
Multiplier test for autoregressive conditional heteroscedasticity. The null
hypothesis in this test is that there is no serial correlation of the
heteroscedasticity. According to the Engle's heteroscedasticity test
results, we can reject the null hypothesis of no serial correlation of the
heteroscedasticity. Hence, we can conclude that there is a serial
correlation of the heteroscedasticity in the mean equation.

\subsection{Estimation results}

In the specified GARCH model, the conditional mean equation is estimated
simultaneously with the conditional variance equation because variance is a
function of the mean. Results for both mean and variance equations are
reported in Table 3.

As regards the \textit{statistical significance} of our results, in the
conditional mean equation most of variables are not significantly different
from zero (Table 3, rows 3-10), which is in line with our expectations, as
all explanatory variables are lagged by one period. If these variables were
statistically significant, it could present an opportunity for arbitrage.
Thus, we can conclude that none of explanatory variables in the previous
period are significant in forecasting the current period's log returns of
BitCoin. In other words, BitCoin's returns are independent from the
influence of all analysed explanatory variables and there is no arbitrage
opportunity.

In the variance equation, we can observe that all explanatory variables are
significantly different from zero at the 99\% confidence interval (rows
12-17 in Table 3). Note that the overall significance of the variance
equation is considerably higher than in the mean equation (rows 3-10 in
Table 3). We can also observe that both ARCH and GARCH terms are
statistically significant (Rows 19 and 20 in Table 3), which implies that
the previous period's return information of BitCoin affects the current
period's volatility of BitCoin (ARCH) and also that the previous period's
volatility of BitCoin influences the current period's volatility of BitCoin
(GARCH).

As regards the \textit{sign} of estimated coefficients, they all are in line
with our expectations from the theoretical model (see equation (3)). BitCoin
returns are increasing in the size of the BitCoin economy (rows 13 and 14 in
Table 3). In contrast, BitCoin returns are decreasing in velocity, the total
BitCoin stock and the general interest rate (rows 12, 15, 16 and 17 in Table
3). Indeed, the estimated coefficients of the total BitCoin stock (\textit{%
logtot\_btc}), velocity (\textit{logvelocity}) and interest rate (\textit{%
logr\_rate}) have negative signs. In contrast, the estimated coefficients of
the traded BitCoin volume (\textit{logvolume}) and the number of BitCoin
users (\textit{logno}) -- which both proxy the size of the BitCoin economy
-- are positive.

In terms of the \textit{magnitude} of the estimated effects, our estimates
suggest that BitCoin returns are more affected by the total BitCoin stock
and the exchange rate. The impact of the number of transactions and velocity
on BitCoin returns is less pronounced. As regards the magnitude of ARCH and
GARCH terms, we can observe that the GARCH coefficient is larger than the
ARCH coefficient (Rows 19 and 20 in Table 3), which implies that past
volatility effects are superior to past shock effects and hence past
volatility effects should be used when forecasting the BitCoin's volatility.

\section{Conclusions}

Both theoretical models and empirical studies have been trying to understand
the mechanics of the virtual currency price formation. Among others,
previous studies have looked at factors related to the blockchain technology
and its implication for financial markets as well as the virtual currency
price formation. Although, there is a growing literature in this field, the
existing evidence is inconclusive in terms of providing a conceptually and
empirically consistent explanation of the BitCoin price development. One
possible causes of inconclusive results is rooted in the underlying data --
the large majority of the existing empirical literature on virtual
currencies is based on rather aggregated (either on daily or weekly) data,
which however masks a great deal of complexity surrounding the virtual
currency price formation.

The present study attempts to shed additional light on the highly complex
BitCoin price formation dynamics by making use of high frequency data. To
our knowledge, this is the first paper that estimates the price determinants
of BitCoin in a GARCH framework using high frequency data.

In order to identify and estimate drivers of the BitCoin price, first we
derive a conceptual model of the BitCoin price formation. In a second step,
building on previous empirical studies on the BitCoin price formation, we
apply a GARCH model to estimate factors affecting the BitCoin price using
hourly data for the period 2013--2018.

Our empirical results confirm that the BitCoin transaction demand and
speculative demand have a statistically significant impact on the BitCoin
price formation. The BitCoin price responds negatively to BitCoin velocity,
whereas positive shocks to the BitCoin stock, interest rate and the size of
the BitCoin economy exercise an upward pressure on the BitCoin price. The
high frequency (hourly) data analysed in the present study allow to gain
additional insights, which remain masked using averaged daily or weekly
prices. Our results suggest that this is a promising avenue for future
research and should be pursued also for other virtual currencies.

\clearpage\newpage

\newpage
\includegraphics[scale=0.8,angle=90,origin=c,trim=1cm 1cm 1.5cm 1cm,
clip=true]{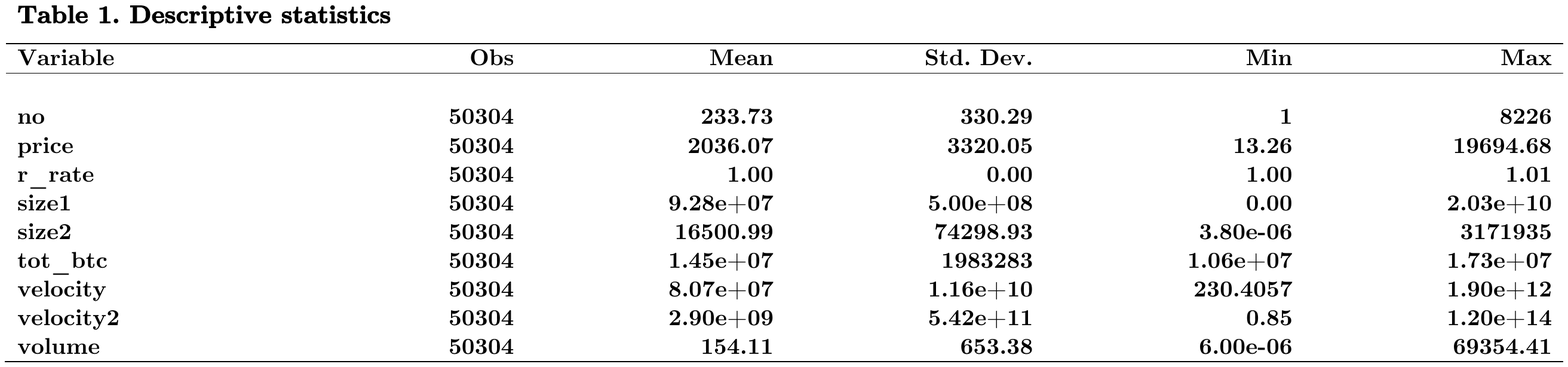} \newpage
\includegraphics[scale=0.8,angle=90,origin=c,trim=1cm 1cm 1.5cm 1cm,
clip=true]{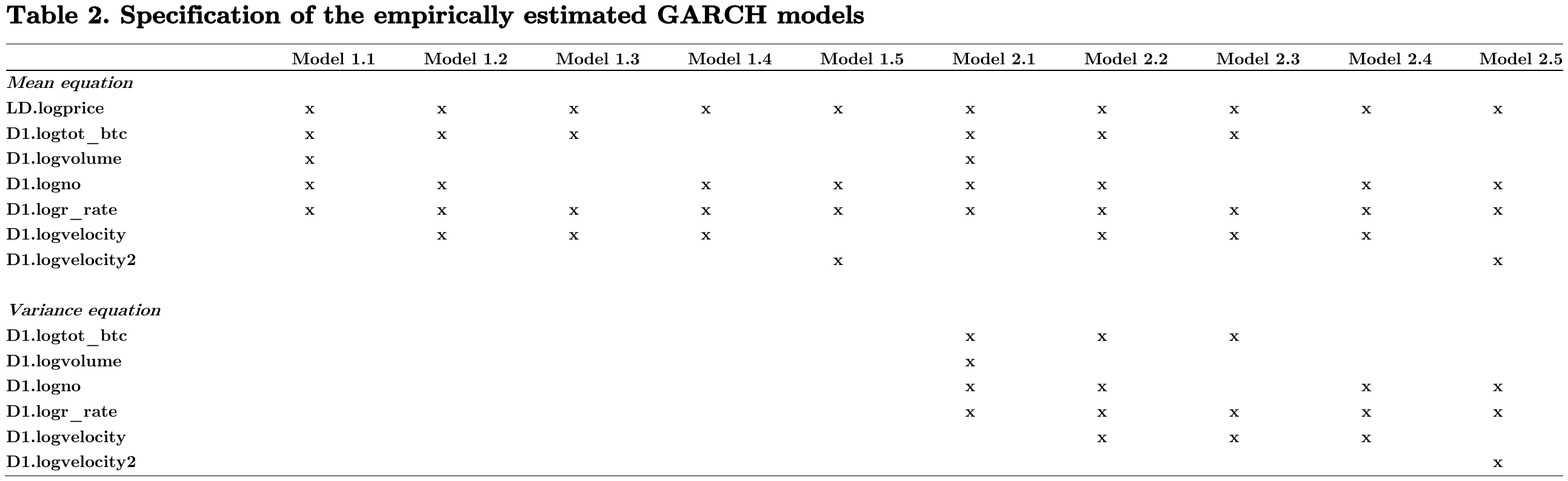} \newpage
\includegraphics[scale=0.8,angle=90,origin=c,trim=1cm 1cm 1.5cm 1cm,
clip=true]{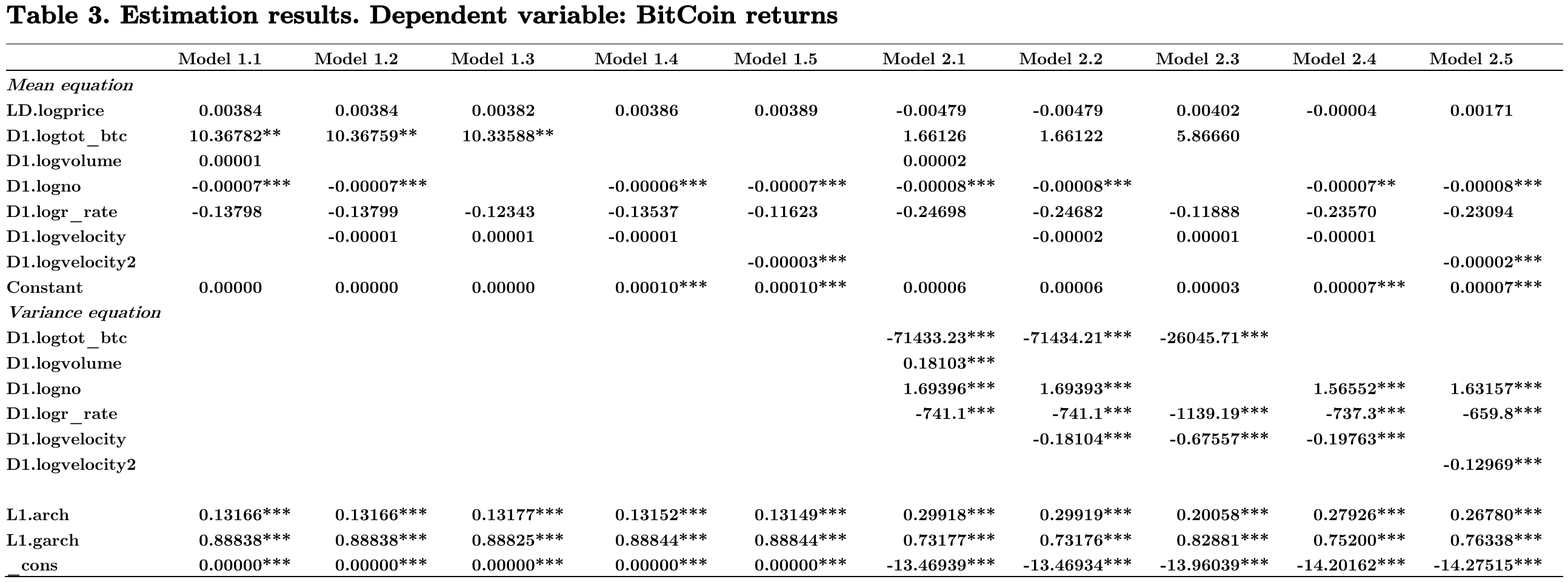} \newpage

\begin{figure}[htbp]
\centering
\includegraphics[scale=1]{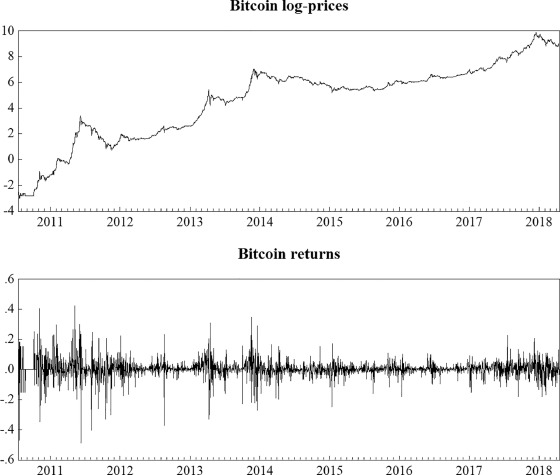}
\caption{BitCoin price (top panel) and BitCoin returns (bottom panel)}
\label{fig:BTC_price_return}
\end{figure}


\section{References}

\begin{thebibliography}{99}
\bibitem{ Aalborg, H.A., P. Moln} Aalborg, H.A., P. Moln\'{a}r and J.E. de
Vries (2018) What can explain the price, volatility and trading volume of
BitCoin? Finance Research Letters Forthcoming.

\bibitem{ Barber, S., X. Boyen,} Barber, S., X. Boyen, E. Shi, and E. Uzun
(2012). Bitter to Better-How to Make BitCoin a Better Currency. In Financial
Cryptography and Data Security. Vol. 7397, of Lecture Notes in Computer
Science, edited by, Keromytis, A. D., 399-414. Berlin: Springer.

\bibitem{ Baur, D.G., K. Hong an} Baur, D.G., K. Hong and A.D. Lee (2018).
BitCoin: Medium of exchange or speculative assets? Journal of International
Financial Markets, Institutions and Money 54: 177-189.

\bibitem{ Bollerslev, T. (1986).} Bollerslev, T. (1986). Generalized
autoregressive conditional heteroskedasticity. Journal of Econometrics 31:
307-27.

\bibitem{ Bouoiyour 2015} Bouoiyour, J., and R. Selmi (2015). What Does
BitCoin Look Like? MPRA Paper No. 58091. Germany: University Library of
Munich.

\bibitem{ Bouoiyour, J., R. Selm} Bouoiyour, J., R. Selmi, A.K. Tiwari, O.R.
Olayeni (2016). What drives BitCoin price? Economics Bulletin, 36(2): 1-9.

\bibitem{ Bouri, E., P. Molnár,} Bouri, E., P. Moln\'{a}r, G. Azzi, D.
Roubaud and L.I. Hagfor (2017). On the hedge and safe haven properties of
BitCoin: Is it really more than a diversifier? Finance Research Letters 20:
192-198.

\bibitem{ Brooks, Ch. (2014). "I} Brooks, Ch. (2014). Introductory
econometrics for finance. Cambridge university press.

\bibitem{ Brownlees, C.T., Engle} Brownlees, C.T., Engle, R.F., Kelly, B.T.
(2011). A practical guide to volatility forecasting through calm and storm
(August 1, 2011). \url{http://dx.doi.org/10.2139/ssrn.1502915}

\bibitem{ Buchholz, M., J. Delan} Buchholz, M., J. Delaney, J. Warren, and
J. Parker (2012). Bits and Bets, Information, Price Volatility, and Demand
for BitCoin, Economics 312. \url{www.bitcointrading.com/pdf/bitsandbets.pdf.}

\bibitem{ Cermak, V. (2017). Can} Cermak, V. (2017). Can BitCoin become a
viable alternative to fiat currencies? An empirical analysis of BitCoin's
volatility based on a GARCH model. \url{https://ssrn.com/abstract=2961405}

\bibitem{ Chen, S. et al., (2016} Chen, S. et al., (2016). A first
econometric analysis of the CRIX family. %
\url{https://ssrn.com/abstract=2832099}

\bibitem{ Chu,J., Chan, S., Nada} Chu,J., Chan, S., Nadarajah, S. and J.
Osterrieder. (2017). GARCH Modelling of Cryptocurrencies, Journal of Risk
and Financial Management, 10(4), 1-15.

\bibitem{ Ciaian Rajcani2016} Ciaian, P., M. Rajcaniova, D. Kancs (2016).
The economics of BitCoin price formation. Applied Economics, 48 (19),
1799-1815.

\bibitem{ Ciaian Rajcani2018} Ciaian, P., M. Rajcaniova, D. Kancs (2018).
Virtual relationships: Short- and long-run evidence from BitCoin and altcoin
markets. Journal of International Financial Markets 52: 173-195.

\bibitem{ Dyhrberg, A.H. (2016a)} Dyhrberg, A.H. (2016a). Hedging
capabilities of BitCoin. Is it the virtual gold? Finance Research Letters
16: 139-44.

\bibitem{ Dyhrberg, A.H. (2016b)} Dyhrberg, A.H. (2016b). BitCoin, gold and
the dollar--A GARCH volatility analysis. Finance Research Letters 16:
85-92.

\bibitem{ Dwyer, P. Gerald 2015,} Dwyer, P. Gerald 2015, The economics of
BitCoin and similar private digital currencies, Journal of Financial
Stability, 17, 81-91.

\bibitem{ Engle, R.F. (1982) "Au} Engle, R.F. (1982) Autoregressive
conditional heteroscedasticity with estimates of the variance of United
Kingdom inflation. Econometrica, 987-1007.

\bibitem{ Fisher I. (1911), The} Fisher I. (1911), The Purchasing Power of
Money.

\bibitem{ Folkinshteyn D, Lennon} Folkinshteyn D, Lennon M, Reilly T (2015)
The BitCoin mirage: an oasis of financial remittance. Journal of
International Studies, 10:118-124.

\bibitem{ Gandal, N., J.T. Hamri} Gandal, N., J.T. Hamrick, T. Moore, T.
Oberman (2018). Price manipulation in the BitCoin ecosystem. Journal of
Monetary Economics 95: 86-96.

\bibitem{ Grinberg, R. (2011). "} Grinberg, R. (2011). BitCoin: An
Innovative Alternative Digital Currency. Hastings Science and Technology Law
Journal 4: 159-208.

\bibitem{ Gronwald, M. (2014). "} Gronwald, M. (2014). The economics of
bitcoins market characteristics and price jumps. (December 29, 2014). CESifo
Working Paper Series No. 5121. \url{https://ssrn.com/abstract=2548999}

\bibitem{ Hansen, P.R., Lunde, A} Hansen, P.R., Lunde, A. (2005). A forecast
comparison of volatility models: does anything beat a garch (1, 1)? Journal
of applied econometrics, 20, 873-889.

\bibitem{ Howden, D. (2013). "Th} Howden, D. (2013). The Quantity Theory of
Money. Journal of Prices and Markets 1.1: 17-30.

\bibitem{ Jang, H. and Lee J. (2} Jang, H. and Lee J. (2018). An Empirical
Study on Modeling and Prediction of BitCoin Prices With Bayesian Neural
Networks Based on Blockchain Information. IEEE Access 6: 5427-5437.

\bibitem{ Kalev, P., Liu, W., Ph} Kalev, P., Liu, W., Pham, P. K. and E.
Jarnecic (2004). Public Information Arrival and Volatility of Intraday Stock
Returns. Journal of Banking and Finance, 28, 1441-1467

\bibitem{ Katsiampa, P. (2017).} Katsiampa, P. (2017). Volatility estimation
for BitCoin: A comparison of GARCH models. Economics Letters. 158: 3-6.

\bibitem{ Keynes, J.M., 1936. Th} Keynes, J.M., 1936. The General Theory of
Employment, Interest and Money. Macmillan, London.

\bibitem{ Kristoufek, L. (2013).} Kristoufek, L. (2013). BitCoin Meets
Google Trends and Wikipedia: Quantifying the Relationship between Phenomena
of the Internet Era. Scientific Reports 3 (3415): 1-7.

\bibitem{ Letra, I.J.S. (2016).} Letra, I.J.S. (2016). What drives
cryptocurrency value? A volatility and predictability analysis. %
\url{https://www.repository.utl.pt/handle/10400.5/12556}

\bibitem{ Mankiw NG (2007) Macro} Mankiw N.G. (2007) Macroeconomics, 6th
edn. Worth Publishers, New York.

\bibitem{ Moore, T., and N. Chri} Moore, T., and N. Christin. 2013. Beware
the Middleman: Empirical Analysis of BitCoin-Exchange Risk. Financial
Cryptography and Data Security 7859: 25-33.

\bibitem{ Poon, S.H., Granger, C} Poon, S.H., Granger, C. (2005). Practical
issues in forecasting volatility. Financial Analysts Journal, 61, 45-56.

\bibitem{ Stavroyiannis, S. and} Stavroyiannis, S. and V. Babalos. (2017).
Dynamic properties of the BitCoin and the US market. %
\url{https://ssrn.com/abstract=2966998}

\bibitem{ Urquhart, A. (2017). "} Urquhart, A. (2017). The volatility of
BitCoin. \url{https://ssrn.com/abstract=2921082}

\bibitem{ van Wijk, D. (2013). W} van Wijk, D. (2013). What can be expected
from the BitCoin? Working Paper No. 345986. Rotterdam: Erasmus Rotterdam
Universiteit.

\bibitem{ Vlastakis, N. and R.N.} Vlastakis, N. and R.N. Markellos. (2012).
Information demand and stock market volatility. Journal of Banking and
Finance, 36, 1808-1821.
\end{thebibliography}
\end{document}